\documentstyle[11pt]{article}
\input epsf.sty
\newcommand{\beq}{\begin{equation}}
\newcommand{\eeq}{\end{equation}}
\newcommand{\beqa}{\begin{eqnarray}}
\newcommand{\eeqa}{\end{eqnarray}}

\title{The Phase Structure of an $SU(N)$ Gauge Theory with $N_f$ Flavors}
\author{Thomas Appelquist \\
Department of Physics, Yale University, New Haven, CT 06511
\\ \\
Anuradha Ratnaweera\\
Institute of Fundamental Studies, Kandy, Sri Lanka
\\  \\
John Terning \\
Department of Physics, University of California\\
Berkeley, CA 94720\\
{\it and}\\
Theory Group, Lawrence Berkeley National Laboratory\\
Berkeley, CA 94720
\\  \\
L.C.R. Wijewardhana \\
Department of Physics, University of Cincinnati, Cincinnati,
OH 45221}
\begin{document}
\setlength{\baselineskip}{24pt}
\maketitle
\begin{picture}(0,0)(0,0)
\put(295,455){YCTP-P15-98}
\put(295,444){UCB-PTH-98/34}
\put(295,433){LBNL-41946}
\put(295,422){UCTP-110-98}
\end{picture}
\vspace{-44pt}

\begin{abstract}
We investigate the chiral phase transition in $SU(N)$ gauge
theories as the number of quark flavors, $N_f$, is varied.
We argue that the transition takes place at a large enough value of $N_f$
so that
it is governed by the infrared fixed point of the $\beta$ function.
We study the
nature of the phase transition analytically and numerically, and discuss
the spectrum of the theory as the critical value of $N_f$ is approached in
both the symmetric
and broken phases. Since the transition is governed by a conformal
fixed point, there are no light excitations on the symmetric
side. We extend previous work to include higher order effects by developing
a renormalization group estimate of the critical coupling.
\end{abstract}

\section{Introduction}

     In an $SU(N)$ gauge theory with $N_f$ massless quarks, it is
expected that both confinement and spontaneous chiral symmetry breaking
take place providing that $N_f$ is not too large.
If, on the
other hand, $N_f$ is large enough, the theory is expected to neither confine
nor break chiral symmetry. For example, if $N_f$ is larger than $11N/2$ for
quarks in the fundamental representation, asymptotic freedom (and hence
confinement and chiral symmetry breaking) is lost. Even for a range of $N_f$
below  $11N/2$, the theory should remain chirally symmetric and deconfined. The
reason is that an infrared fixed point is present
\cite{GW,CasJones} determined by
the first two terms in the renormalization group (RG) beta function. By an
appropriate choice of $N$ and $N_f$, the coupling at the fixed point,
$\alpha_*$, can be made arbitrarily small \cite{BZ}, making a perturbative
analysis
reliable. Such a theory is massless and conformally invariant in the infrared.
It is asymptotically free, but without confinement or chiral symmetry
breaking.

     As $N_f$ is reduced, $\alpha_*$ increases.  At some critical value of
$N_f$ ($N_f^c$)
there will be a phase transition to the chirally asymmetric and confined
phase.
It is an important problem in the study of gauge field theories to
determine $N_f^c$ and to characterize the nature of the phase
transition.

In a recent letter \cite{ATWFP}, we suggested that the phase transition
takes place at a large enough value of $N_{f}^c$ so that the infrared fixed
point
$\alpha_*$ reliably exists and governs the phase transition. The transition
was then analyzed
using the ladder expansion of a gap equation, or equivalently the CJT
effective potential \cite{CJT}. It was argued that confinement  effects can
be neglected to estimate $N_f^c$ and to
determine the nature of the transition. It was then shown that the chiral
order parameter  vanishes continuously at  $ N_f  \rightarrow N_f^c$ from
below, but that the phase transition is not conventionally second order in
that there is no effective, low energy Landau-Ginzburg Lagrangian, i.e.
the correlation length does not diverge as the critical point is approached.

  Once chiral symmetry breaking sets in, the quarks decouple at momentum
scales below the dynamical mass leaving the pure gauge theory behind. The
effective coupling then grows, leading to confinement at a scale on the
order of the quark mass. Thus for $N_f$ just below $N_f^c$, the fixed
point is only an approximate feature of
the theory governing momentum scales
above the dynamically generated mass. This is adequate, however, since it
is this momentum range that determines $N_f^c$ and the character of the
transition.

     Our discussion of this phase transition paralleled an analysis of
the chiral transition in $2+1$ dimensional gauge theories with
$N_f$ quarks \cite {ATW2+1}.
Using a large $N_f$ expansion it was found \cite{QED3} that the
effective  infrared coupling  runs to a fixed point proportional to
$1/N_f$.  As
$N_f $ is lowered this coupling strength exceeds the critical coupling
necessary to produce
spontaneous symmetry breaking.  It was argued that this critical
$1/N_f$ coupling lies in a range where the large $N_f$ expansion is
reliable \cite{Nash}. These conclusions were
also supported by lattice simulations
\cite{Kogut}. It was then noted that as in the case of the 3+1
dimensional $SU(N)$ theory, this phase transition is not
conventionally second order \cite {ATW2+1}.

 For QCD the study of the chiral phase transition as a function
of $N_f$ is of theoretical interest, but
 is unlikely to shed direct light on the physics of the real world. There
remains the possibility, however, that if technicolor is the correct
framework for electroweak symmetry breaking,
the transition could be physically relevant. In a recent letter
\cite{postmodern}, it was pointed out that in an $SU(2)$
technicolor theory, a single family of techniquarks ($N_f = 8$) leads to
an infrared fixed point near the critical coupling for
the chiral phase transition. This can provide a natural origin \cite{LaneRam}
for walking technicolor \cite{walking} and has other interesting
phenomenological features.

In this paper, we explore further the features of the chiral phase
transition as function of $N_f$.
 In Section 2, we summarize the properties of an $SU(N)$ gauge theory with
$N_f$
massless quarks,
and describe the existence and properties of
an infrared (IR) stable fixed point.
In Section 3, we review chiral phase transition lore in $SU(N)$ gauge
theories, both at zero temperature and finite temperature.
We present our  study of the chiral phase transition in Section 4. We
examine the character of
the phase transition by
computing the quark-antiquark scattering amplitude for $N_f > N_f^c$
($\alpha_* < \alpha_c$) in the RG improved ladder approximation. We observe
that for
$\alpha_* \rightarrow
\alpha_c$ from below, there are no light scalar or pseudo-scalar degrees of
freedom,
showing that
the phase transition is not conventionally second order. A light spectrum,
in addition to the Goldstone bosons, does exist in the broken phase, and we
describe what is currently known about it.  In section 5, we include the
effects
of higher order contributions to both the
RG $\beta$ function and the estimate of the critical coupling,
and then discuss the reliability of our results.
In Section 6, we summarize our results, compare them to those from other
recent studies of $SU(N)$ theories, and make some comparisons of our work to
the phase structure of
supersymmetric gauge theories. In an appendix, we discuss infrared and
collinear divergences, and issues of gauge invariance arising in the study
of the quark-antiquark scattering amplitude.

\section{Features of an $SU(N)$ Gauge Theory with $N_f$ Flavors}

The Lagrangian of an $SU(N)$ gauge theory is:
\beq
 {\cal L} = \bar{\psi}(i\not\!\partial + g(\mu)\not\!\!A^a T^a)\psi
-
{1 \over 4} F^a_{\mu\nu}F^{a\mu\nu}
\label{L}
\eeq
where $\psi$ is a set of $N_f$ 4-component spinors, the $T^a$ are
the generators of $SU(N)$, and $g(\mu)$
is the gauge coupling
defined by integrating out momentum components above $\mu$.
With no quark mass, the quantum theory is invariant under the global
symmetry group $SU(N_f)_L \times SU(N_f)_R \times U(1)_{L+R}$.

The RG equation for the running gauge coupling  is:
\beq
\mu{{\partial}\over{\partial \mu}} \alpha(\mu) = \beta(\alpha)
\equiv -b\, \alpha^2(\mu) -c\, \alpha^3(\mu)-d\, \alpha^4(\mu) - ...~,
\label{beta}
\eeq
where $\alpha(\mu) = g^2(\mu)/4 \pi$.  With $N_f$ flavors of quarks in the
fundamental representation, the first two coefficients are given by
\beq
b = {{1}\over{6 \pi}} \left( 11 N - 2 N_f\right)
\label{b}
\eeq
\beq
c = {{1}\over {24  \pi^2}} \left(34 N^2 - 10  N N_f - 3{{N^2 -
1}\over{N}} N_f\right)~.
\label{c}
\eeq
These two coefficients are independent of the renormalization scheme. The
theory is asymptotically free if $b > 0$ ($N_f < {{11}\over{2}}N$).
At two loops, the theory has
an infrared stable, non-trivial fixed point if $b > 0$ and $c < 0$.  In this
case the fixed point
is at
\beq
\alpha_* = - \,{{ b}\over {c}}~.
\eeq

The fixed point coupling $\alpha_*$ can be made arbitrarily small by taking
$(11 N/2 -
N_f)/N$ to be small and positive \cite{BZ}.
This can be achieved either by going to large $N$ and $N_f$ with the ratio
fixed, or by analytically continuing in $N_f$.  With the coupling taken to
run between zero in the ultraviolet and $\alpha_*$ in the infrared, the
higher order terms in $\beta(\alpha)$ can then reliably be neglected. The
theory is only weakly interacting in the infrared, so that there is no
chiral symmetry breaking or confinement.

At two-loops the solution of the RG equation can be written as:
\beq
b\log\left({{q}\over{\mu}}\right) = {{1}\over{\alpha}} -
{{1}\over{\alpha(\mu)}}- {{1}\over{\alpha_*}}\log\left({{\alpha
\left(\alpha(\mu) -
\alpha_*\right)}\over{\alpha(\mu)\left(\alpha-\alpha_*\right)}}\right)~,
\eeq
where $\alpha = \alpha(q)$.  For $\alpha$, $\alpha(\mu)<\alpha_*$ we can
introduce a scale defined by
\beq
\Lambda = \mu \exp\left[{{-1}\over{b \,\alpha_*}}
\log\left({{\alpha_*-\alpha(\mu)}
\over{\alpha(\mu)}}\right)-{{1}\over{b \,\alpha(\mu)}}\right]~,
\label{Lambda}
\eeq
so that
\beq
{{1}\over{\alpha}} = b \log\left({{q}\over{\Lambda}}\right) +
{{1}\over{\alpha_*}}
\log\left({{\alpha}\over{\alpha_*-\alpha}}\right).
\eeq
Then for $q \gg \Lambda$ the running coupling displays the usual
perturbative behavior:
\beq
\alpha \approx {{1}\over{b \log\left({{q}\over{\Lambda}}\right)}}~,
\label{highalpha}
\eeq
while for $q \ll \Lambda$ it approaches the fixed point $\alpha_*$:
\beq
\alpha \approx {{\alpha_*}\over{1+ {{1}\over{e}}
\left({{q}\over{\Lambda}}\right)^{b \alpha_*}}}~.
\label{lowalpha}
\eeq
Thus for $N_f$ in the range where an infrared fixed-point exists, $\Lambda$
represents the intrinsic scale of the theory: above the scale $\Lambda$ the
coupling becomes asymptotically free, while below $\Lambda$ the
coupling rapidly approaches the infrared fixed-point.

It is interesting to note that the solution for $\alpha = \alpha(q)$ can be
written generally as
\beq
\alpha = \alpha_* \left[W
  \left(
  q^{b\alpha_*}/e\Lambda^{b\alpha_*} \right)
  +1
\right]^{-1},
\eeq
where
$W(x) = F^{-1}(x)$,
with $F(x) = x e^x$,  is the Lambert W function \cite{Lambert},
\cite{Karliner} .
In the limit of small x, $ W(x) \approx x $,
giving Eq. (\ref{lowalpha}) for $q \ll \Lambda$. In the limit of large x,
$W(x)  \approx  \log x$, giving Eq. (\ref{highalpha})
for $ q \gg \Lambda$.

 \section{Chiral Symmetry Breaking}

    The physics of an $SU(N)$ gauge theory, even at zero temperature,
depends strongly on the number of massless flavors.
As we have just noted, if $(11 N/2 - N_f)/N$ is small, the coupling remains
small at all scales and
the theory neither confines nor spontaneously breaks chiral symmetry. The
quarks and gluons remain massless and the theory
 is governed by an infrared fixed point and is therefore conformally
invariant in the infrared.

    For  $N_f$ small compared to $11N/2$, the situation is quite different.
With $N_f = 0$, lattice simulations indicate that
 the theory confines producing a physical spectrum of massive glueballs. In
the case of real-world QCD ($N = 3$ with two light
flavors), confinement and the spontaneous breakdown of the chiral symmetry
from $SU(2)_L \times SU(2)_R \times U(1)_{L+R}$ to
$SU(2)_{L+R} \times U(1)_{L+R}$ are approximate experimental features, seen
also in lattice simulations. Small $N_f$ can also be
explored by taking the large $N$ limit with $N_f$ fixed. There the chiral
symmetry is $U(N_{f})_L \times U(N_{f})_R$, the
chromodynamic anomaly being irrelevant to leading order. It was was shown by
Coleman and Witten \cite{colemanwitten} that under
reasonable assumptions, confinement then necessarily implies the
spontaneous breaking of  $U(N_f)_L \times U(N_f)_R$ to
$U(N_f)_{L+R}$.

    These two different phases of a zero-temperature  $SU(N)$ theory can be
characterized by a simple chiral order
 parameter,  the expectation value of the quark bilinear
\begin{equation}
M^i_j = \, \langle \bar{q}^i_L q^j_R \rangle,
\end{equation}
a.k.a. the quark condensate.
For some range of  $(11 N/2 - N_f)/N$ small, the order parameter vanishes,
while for $N_f$ small compared to $11N/2$,
 it is non-vanishing. The location and character of the transition
constitute an important and unresolved problem in the study of
gauge field theories. This problem has been studied by the continuum gap
equation method, by the consideration of instanton
configurations, and by lattice simulations. After summarizing the results
of the first approach here, we will comment on the other
approaches and compare the results.

    It is also interesting to compare this phase transition with the finite
temperature transition of an $SU(N)$ gauge theory.
There, the transition is known to be second order \cite{FT2} for $N_f = 2$
and has been argued to be strongly first order \cite{FT1} for
$N_f \geq 3$. An important distinction between finite and zero
temperature is that at finite temperature, the
quarks are screened at  distance scales large compared to the inverse
temperature. This is because in Euclidean field theory
at finite temperature, the integral over the energy is replaced by a sum
over Matsubara frequencies given by $2n\pi \, T$ for bosons
and $(2n+1) \pi \, T$ for fermions, where $n$ is an integer. Only the $n=0$
bosons survive at large distances. Thus to characterize a
finite temperature transition in which the order parameter vanishes
continuously, it isn't necessary to consider the quarks or
fermionic bound states of quarks. This is not the case in the
zero-temperature transition to be considered here. Furthermore,
at zero temperature quarks experience long range interactions,
which are screened  at finite temperature.  These differences have
important consequences.

\section{The Gap Equation with an Infrared Fixed Point}

    We examine the chiral phase transition by making a set of simple
assumptions whose
validity we will examine later. First of all, we assume that the transition
takes place at a value of $N_f$ such that the
 infrared coupling is reliably governed by the two-loop fixed point
described above. Even though this may not be a very small
coupling, we assume that the transition may be studied by focusing on the
underlying quark and gluon degrees of freedom, ignoring
other bound states or resonances that might be formed. Next we assume that
the transition is governed to first approximation by a
gap equation in RG-improved ladder approximation. The most attractive
channel  then corresponds to the breaking pattern $SU(N_f)_L
\times SU(N_f)_R \times U(1)_{L+R}$ to $SU(N_f)_{L+R} \times U(1)_{L+R}$.

   In the broken phase, a common dynamical mass $\Sigma(p)$, with $p$ the
magnitude of a Euclidean momentum,  will then
 be generated for all the $N_f$ quarks. It can be taken to serve as the
order parameter for the chiral phase transition, and is
related simply to the quark condensate. Although this quantity, unlike the
quark condensate, is gauge dependent, it is possible to
extract gauge-independent information from  it.

With only the quark and gluon degrees of freedom employed, an analysis of
the gap equation leads to the conclusion that the chiral transition is one
in which the order parameter vanishes
continuously at the transition.
Near the transition, $\Sigma(p)$ is small compared to the
intrinsic scale $\Lambda$, and the equation can be
linearized to study the momentum regime $ \Sigma(p) <p < \Lambda$ that
dominates the transition. At low momenta the running
coupling $\alpha(k)$ appearing in the gap equation approaches its fixed
point value
$\alpha_*$. It is well known that the gap equation has non-vanishing
solution only when this coupling exceeds a gauge-invariant
critical\footnote{A more general definition \cite{rainbow} of the
critical coupling is that the anomalous dimension of
$\overline{\psi}\psi$ becomes 1.}  value
\beq
\alpha_c \equiv {{ \pi }\over{3 \, C_2(R)}}=
{{2 \pi \,N}\over{3\left(N^2-1\right)}}~.
\label{alphacrit}
\eeq
It can be shown that when the coupling exceeds this critical value, the CJT
effective potential \cite{CJT} becomes unstable at the origin, indicating
that a chirally-asymmetric solution is energetically favored and
therefore represents the ground state of the theory.

Setting $\alpha_*$ equal to $\alpha_c$ gives an estimate \cite{ATWFP} of
the critical number of flavors
\beq
N_f^c = N \left({{100N^2 -66}\over{25 N^2 -15}}\right),
\label{Ncrit}
\eeq
above which there is no chiral symmetry breaking. Note that the ratio $N_f
^c/N$ is predicted to be very close to $4$ for all $N$.

We next discuss the critical behavior at this transition. Since the
infrared behavior is governed by the fixed point $\alpha_*$, we can get a
simplified look at the transition by taking the
coupling to be constant  and equal to $\alpha_* > \alpha_c$ in a momentum
range up to some cutoff $\Lambda_* < \Lambda$.  The well-known solution to
this simplified model (often referred to in the literature as quenched
QED) is a non-vanishing
dynamical mass $\Sigma(p)$ falling monotonically as a
function of $p$ from some value $\Sigma(0)$ \cite{classicrefs,SD}.  For
$\alpha_{*} \rightarrow \alpha_c$ from above  ($N_f  \rightarrow N_f^c$
from below),
$\Sigma(0)$ exhibits the behavior
\beq
\Sigma(0) \approx \Lambda_*
\exp\left({{- \pi}\over{\sqrt{{{\alpha_*}\over{\alpha_c}} -1}}}\right) ~.
\label{critical}
\eeq
Thus the order parameter $\Sigma(0)$ is predicted to vanish
non-analytically as $\alpha_{*} \rightarrow \alpha_c$.

We expect a similar critical behavior in the full theory. After all, the
intrinsic scale
$\Lambda$ introduced in Eq. (\ref{Lambda}), where $\alpha(\Lambda) \approx
0.78 \,\alpha_*$, plays the role of
an ultraviolet cutoff. Asymptotic freedom sets in beyond this scale and the
dynamical mass function falls rapidly ($\sim 1/p^2$). Indeed we find that with
a
running coupling the
critical behavior is exponential as above, but that the coefficient
in the exponential depends on the details of physics at scales on
the order of $\Lambda$. It is not universally $-\pi$.

This can be understood
analytically  in the following manner. Following Ref. \cite{TABrazil},
the gap equation can be converted to differential form with appropriate
boundary conditions, and the solution to the linearized equation
 can be written as
\beq
 \Sigma(p)  = {c \Sigma(0)^2\over p } \sin{ \int^{p}_{a\Sigma(0)} { dk\over
k}\sqrt{\alpha(k)/\alpha_c-1}}
\eeq
for momenta $p$ below the scale $\Lambda_c$ at which
$\alpha(\Lambda_c)= \alpha_c$, where $c$ is chosen so that
$\Sigma(\Sigma(0)) = \Sigma(0)$.
We have dropped terms explicitly proportional to derivatives of $\alpha(k)$
since the coupling is near the fixed point in this range and we have taken the
lower limit of the integral to be of order $\Sigma(0)$ ($a = {\cal O}(1) $).
For $
k > \Lambda_c$,
the solution takes a different form, expressible in terms of a hyperbolic
sine function
when the running is slow. The two solutions must match at $ p = \Lambda_c$
and the upper solution must satisfy the ultraviolet boundary condition.
Note that $\Lambda_c / \Lambda$
vanishes like $(r - 1)^{1/b\alpha_*}$ as $r \rightarrow 1$,
where  $r \equiv \alpha_* /  \alpha_c$.

The matching condition at $\Lambda_c$ says simply that
\beq
 \int^{\Lambda_c}_{a\Sigma(0)} { dk\over
k}\sqrt{\alpha(k)/\alpha_c-1}
\eeq
takes on some value depending on the details of the upper solution. It can
be seen to be finite in the limit $r \rightarrow 1$ and it must be less
than $\pi$
if the dynamical mass is to remain positive for all momenta. (Solutions
with nodes also exist,
but a computation of the vacuum energy \cite{CJT,Bardeen} indicates that the
 nodeless solution represents the stable ground state.) Because $\alpha(k)
\approx \alpha_*$ for small momenta, it can then be seen that $1/ \log (
\Lambda_c / \Sigma(0))$ vanishes like $ \sqrt{r-1}$ as $r \rightarrow 1$.
Since $\Lambda_c / \Lambda$ behaves like $(r - 1)^{1/b\alpha_*}$, it
follows that  $1/ \log (\Lambda / \Sigma(0))$ also vanishes like $
\sqrt{r-1}$ as $r \rightarrow 1$.

This can also be seen in a direct, numerical solution of the integral gap
equation. In Landau gauge and after Wick rotation to Euclidean space, this
equation can be written in the form
\begin{eqnarray}
\ \Sigma(p) &=&
{1\over4}\int  {{dk^2}\over{M^2}}  {k^2\Sigma(k)\over k^2+\Sigma(k)^2 }{
\alpha(M^2)\over{\alpha_c}}
\end{eqnarray}
where $M = \max(p,k)$ and the approximation $\alpha((p-k)^2) \approx
\alpha(M^2)$
has been made before doing the angular integration. We solve this equation
with a numerical ultraviolet cutoff
much larger than $\Lambda$ and plot  $\log (\Sigma(0)/\Lambda_c)$ versus
$  1 /  \sqrt{r-1}$ in Figure 1. The
result is insensitive to the numerical cutoff and
exhibits straight line behavior as $r \rightarrow1$. The slope of the line
is $0.82 \pi$.
If the theory is modified in some way at scales on the order of $\Lambda$,
straight line behavior is still exhibited, but with a slope depending
on the details of the modification.
Thus the only feature of the critical behavior determined purely by the
infrared,
fixed point  behavior  is that $1/ \log (\Lambda / \Sigma(0))$ vanishes
like $ \sqrt{r-1}$ as $ r \rightarrow 1$.

Below the scale of  the dynamical mass $\Sigma(p)$, the quarks
decouple, leaving a pure gauge theory behind. One might worry that this would
invalidate the above analysis since it relies on the fixed point
which only exists when the quarks contribute to the $\beta$ function.  This
is not a problem, however, since when
$\Sigma(0) \ll \Lambda$, the dominant momentum
range in the gap equation, leading to the above critical behavior
(\ref{critical}), is $\Sigma(0) < p < \Lambda$. In this range, the quarks are
effectively massless and the coupling does appear to be approaching an infrared
fixed point.  Below the scale $\Sigma(0)$ confinement sets in. The
confinement scale can be estimated by noting that
 at the decoupling scale $\Sigma(0)$, the effective coupling constant is of
order $\alpha_c$. A simple estimate using the above
expressions then shows that the confinement scale is roughly the same order
as the chiral symmetry breaking scale, $\Sigma(0)$.

If $N_f $ is
reduced sufficiently below $N_{f}^c$ so that $\alpha_*$ is not close to
$\alpha_c$,  both $\Sigma(0)$ and the confinement scale
become of order $\Lambda$. The linear approximation to the gap equation
is then no longer valid and it is no longer the
case that higher order contributions to the effective potential can be
argued to be small. The methods of this paper are then no longer useful.

 From the behavior of $\Sigma(0)$ near the transition, the corresponding
behavior of the Goldstone boson decay constant,
 the quark condensate, and other physical scales can be  estimated. We
return to this question after considering further
the nature of the chiral phase transition we have just described.

  The smooth vanishing of the order parameter $\Sigma(0)$, Eq.
(\ref{critical}), suggests that the  chiral symmetry phase
transition at $N_f = N_{f}^c$ ($\alpha_{*} = \alpha_{c}$) might be second
order. In a second order transition, however, an infinite
correlation length is associated with a set of scalar and pseudoscalar
degrees of freedom, with vanishing masses, described by an
effective Landau-Ginzburg Lagrangian. In the broken phase, the Goldstone
bosons are massless and the other scalar masses vanish at
the transition. There are no other light degrees of freedom. In the
symmetric phase, the scalars and pseudoscalars form a
degenerate multiplet.  The situation here is quite different. We first
demonstrate this by showing that in the symmetric phase, there are no light
scalar and pseudoscalar degrees of freedom. We then
comment more generally on the physics of the transition.

\subsection{The Symmetric Phase}

To search for light, scalar and pseudoscalar degrees of freedom in the
symmetric phase, we examine
the color-singlet quark-antiquark scattering amplitude in the same
(RG-improved ladder) approximation leading to the above
critical behavior. If the transition is second order, then  poles
should appear which move to zero momentum as we approach the
transition. We take the incoming (Euclidean) momentum of the initial
quark and
antiquark to be $q/2$, but
keep a non-zero momentum transfer by assigning outgoing momenta
$q/2 \pm p$ for the final quark
and antiquark. Any light scalar resonances should make their presence
known by producing pole in the scattering amplitude (in the complex
$q^2$ plane).

If the Dirac indices of the initial quark and antiquark are $\lambda$ and
$\rho$, and the those of the final state quark and antiquark are $\sigma$
and $\tau$, then the scattering
amplitude can be written for sufficiently small $q$ as:
\begin{equation}
T_{\lambda \rho \sigma \tau}(p,q) = \delta_{\lambda \rho} \delta_{\sigma
\tau}\,
{{1}\over{p^2}} \,T(p,q) + . . .~,
\label{fullscatt}
\end{equation}
where the dots indicate pseudoscalar, vector, axial-vector, and tensor
components, and we have factored out $1/p^2$ to make $T(p,q)$
dimensionless.   We contract
Dirac indices so that we obtain the Schwinger-Dyson (SD) equation for the
scalar
s-channel scattering amplitude, $T(p,q)$, containing only t-channel
gluon exchanges.  If $p^2 \gg q^2$, then $q^2$ will simply act as an
infrared cutoff in the loop integrations.

The SD equation  in the scalar channel
is:
\begin{eqnarray}
T(p,q) &=& {{\alpha_*}\over{\alpha_c}} \pi^2+
4\pi^2 \lambda \,{{ p^2}\over{\Lambda_*^2}}+
{{\alpha_*} \over{4\alpha_c}}\left(
\int_{q^2}^{p^2} {{dk^2}\over{k^2}} \,T(k,q) +
\int_{p^2}^{\Lambda_*^2} {{dk^2}\over{k^2}} \,T(k,q) \,{{ p^2}\over{k^2}}
\right)
\nonumber \\
& &+
\lambda \int_{q^2}^{\Lambda_*^2} {{dk^2}\over{k^2}} \,T(k,q) \,{{
p^2}\over{\Lambda_*^2}} ~.
\label{T}
\end{eqnarray}
For the purpose of this discussion
we neglect the running of the gauge coupling
$\alpha$ up to the scale $\Lambda_*$.
This is a good approximation at the low momenta of interest here,
where the coupling is near the infrared fixed point $\alpha_*$. For
convenience, we use Landau gauge ($\xi=1$) where the quark
wavefunction renormalization vanishes.  The issue of gauge invariance is
addressed
in the Appendix.
The first term in Eq. (\ref{T}) is simply one gluon
exchange, while the second term arises from a chirally symmetric,
four-quark interaction, i.e. a Nambu---Jona-Lasinio (NJL) \cite{NJL}
interaction, which we have introduced here for purposes of this analysis. It
allows us to make contact with the familiar
 study of light degrees of freedom in the NJL theory when it
is near-critical.

For momenta $p^2>q^2$, Eq.~(\ref{T}) can be converted to a
differential equation:
\begin{equation}
p^4{d^2 \over (dp^2)^2}\,\, T  =
-{{\alpha_*}\over {4\alpha_c}}\,\,T~,
\label{diff}
\end{equation}
with appropriate boundary conditions determined from Eq.~(\ref{T}).
The solutions of Eq.~(\ref{diff}) have the form.
\begin{equation}
T(p,q) = A
\left({p^2 \over \Lambda_*^2}\right)^{{1 \over 2} + {1 \over 2} \eta}
 +B
\left({p^2 \over \Lambda_*^2}\right)^{{1 \over 2} - {1 \over 2} \eta} ~,
\label{sol}
\end{equation}
where the coefficients $A$ and $B$ are functions of  $q^2 / \Lambda_*^2$,
and for $\alpha_*<\alpha_c$,
\begin{equation}
\eta= \sqrt{1-\alpha_*/\alpha_c}~.
\label{eta}
\end{equation}
The coefficients $A$ and $B$ can be
determined by substituting the solution back into Eq.~(\ref{T}).  This
gives:
\begin{equation}
A ={{-2 \pi^2}\over{\left(1+\eta\right)^2}}
{{\left(1-\eta\right) \left(1- {{\lambda}\over{\lambda_*}}\right)
\left(  {{q^2}\over {\Lambda_*^2}}\right)^{-{1\over 2}+{1\over 2}\eta} }
\over { 1- {{\lambda}\over{\lambda_\alpha}}+
\left({{\lambda}\over{\lambda_\alpha}}-
\left({{1-\eta}\over{1+\eta}}\right)^2\right)
\left({{q^2}\over {\Lambda_*^2}}
\right)^\eta }   }
{}~,
\label{A}
\end{equation}
and
\begin{equation}
B =
{{2 \pi^2\left(1-\eta\right)
\left(1- {{\lambda}\over{\lambda_\alpha}}\right)
\left({{q^2}\over {\Lambda_*^2}}\right)^{-{1\over 2}+{1\over 2}\eta} }
\over { 1- {{\lambda}\over{\lambda_\alpha}}+
\left({{\lambda}\over{\lambda_\alpha}}-
\left({{1-\eta}\over{1+\eta}}\right)^2\right)
\left({{q^2}\over {\Lambda_*^2}}
\right)^\eta }   }~,
\label{B}
\end{equation}
where
\begin{equation}
 \lambda_\alpha \equiv \left[{1 \over 2} + {1 \over 2} \eta
\right]^2 ~,
\label{lambda_alpha}
\end{equation}
and
\begin{equation}
\lambda_* \equiv \left[{1 \over 2} - {1 \over 2} \eta
\right]^2 ~.
\label{lambdatilde}
\end{equation}

If we denote the location of the poles of the functions $A$ and $B$ in the
complex $q^2$ plane by $q_0^2$, we
then have
\begin{equation}
|q_0^2| = \Lambda_*^2
\left({| \lambda_\alpha- \lambda |}\over
{|\lambda- \lambda_*|}\right)^{1 \over \eta}~.
\label{q0}
\end{equation}
We see immediately that as $\lambda \rightarrow \lambda_\alpha$ (the
critical NJL coupling)
for $\alpha_* <\alpha_c$ the pole approaches the origin $q_0^2 = 0$,
indicating the existence of light degrees of freedom.
This is to be expected for a second order phase transition. As
$\alpha_*$ is increased the corresponding particles become broad
resonances  \cite{ATW1}.
Of course in this region our analysis is not complete, precisely because
of the existence of the light scalar and  pseudoscalar degrees of freedom.
These light degrees of freedom must be incorporated into the analysis,
for example they will have an effect on the two loop $\beta$ function.
Furthermore as discussed by Chivukula et. al. \cite{criticalhier} one
generally expects that, with more than two flavors of quarks, as
$\lambda$ is tuned towards $\lambda_\alpha$ the theory undergoes a
Coleman-Weinberg transition \cite{colemanweinberg} to the chirally broken
phase before $\lambda$ reaches $\lambda_\alpha$.

Now consider the limit $\eta \rightarrow 0$ ($\alpha_* \rightarrow \alpha_c$),
with
$\lambda < 1/4$, we have
\begin{eqnarray}
|q_0^2| &\rightarrow &
\Lambda_*^2 \left(1+{{\eta}\over{1/4-\lambda}}\right)^{1\over\eta}
\nonumber \\
 &\rightarrow & \Lambda_*^2 \exp\left({{4}\over{1-4\lambda}}\right) ~.
\label{q0limit}
\end{eqnarray}
Thus we see that at $\alpha_* \rightarrow \alpha_c$, with $\lambda<1/4$,
there are no
poles in the complex $q^2$-plane with $q_0^2 \ll\Lambda_*$.  There are
therefore no
light scalar and pseudoscalar degrees of freedom to constitute an effective
Landau-Ginzburg theory, so the chiral phase
 transition is not second order along the line
$\alpha_*=\alpha_c$.  This is in agreement with the analysis of Ref.
\cite{Miransky}.

Now imagine starting out with $\alpha_* < \alpha_c$ and $\lambda \approx
\lambda_\alpha$, so that
we have a light scalar resonance,
and then dialing the parameters so that $\alpha_*$ increases and $\lambda$
decreases in such a way that we approach the critical line $\alpha_*=\alpha_c$.
We then see from Eqs. (\ref{q0}) and (\ref{lambdatilde}) that we must
first
cross the line $\lambda = \lambda_*$, and that as we approach this line,
the mass of the scalar
grows and actually diverges.  Thus the scalar resonance disappears from the
physical spectrum before we reach $\alpha_* = \alpha_c$.
Even before we reach this point, the width of the scalars becomes as large as
their mass, and they can no longer be considered resonances.

There is nothing special about the scalar and pseudoscalar channels in the
above
analysis. A similar analysis of the other channels, such as vector and
axial-vector,
would also reveal that there are no light excitations in the symmetric
phase near the critical coupling $\alpha_c$.
  That this should be the case is not surprising. With the transition
governed by a long-range gauge force with an infrared fixed point,
approximate conformal invariance should be exhibited at momentum scales small
compared to $\Lambda$ in the symmetric phase. (For further discussions on this
point see Ref. \cite{miryam}.) Thus no light
scales will be present, in contrast to phase transitions governed by short
range forces as in the NJL or the finite temperature theories.

\subsection{The Broken Phase}

  In the broken phase near the transition, one light scale, $\Sigma(0)$,
appears. It is therefore natural (in the assumed absence of instanton
effects) to expect
that the entire physical spectrum of the theory will be set by $\Sigma(0)$
and scale to zero with it as $N_f \rightarrow N_f^c$
from below. This point has been stressed recently by Chivukula
\cite{chivukula}. Thus there will clearly be no effective
Landau-Ginzburg Lagrangian. No finite set of light degrees of freedom can
be isolated in the broken phase in the limit  $N_f \rightarrow N_f^c$,
and no light degrees of freedom (other than quarks and gluons)
exist in the symmetric phase!

Within this general picture, it is important to describe the spectrum of
resonances in more detail. If, for example,
a near-critical theory is the basis for a technicolor theory of
electroweak symmetry breaking \cite{postmodern}, then the the
light scale $\Sigma(0)$ will correspond to the electroweak scale and the
spectrum of resonances at this scale will have a direct
impact on precision electroweak measurements. In particular, the $S$
parameter \cite{S} will depend sensitively on this
spectrum. An especially interesting question in this regard is whether
parity doubling or even inversion of parity partners appears in this light
spectrum as
$N_f^c$ is approached.

The Goldstone boson decay constant $F_\pi$ is also proportional to
$\Sigma(0)$. A simple dimensional estimate
suggests that $F_\pi^2 \approx N \Sigma^{2}(0)/16 \pi^{2}$. Because of the
dominance of the fixed point at scales below $\Lambda$, this is
clearly a ``walking" theory.  If the coupling stays close to $\alpha_c$ then
the
dynamical mass $\Sigma(p)$ falls roughly like $1/p$ in
this range. As a consequence,  the condensate
$\langle \bar{q}^i_L q^j_R \rangle$ is
enhanced well above the value it would have in a QCD-like
theory. A simple estimate gives $< \bar{q}^i_L q^j_R > \approx N
\Sigma(0)^2 \Lambda /16 \pi^2 $.

 Finally, it is important to note that with the entire spectrum of physical
states collapsing to zero with $\Sigma(0)$ at the transition, the analysis
of the transition using only the quark and gluon degrees of freedom is open
to question. It seems reasonable, however, to conjecture that these states
will
not be important at the momentum scales $ \Sigma(0) < k < \Lambda$
dominating the transition.
Some evidence for this is provided by estimates of higher order effects to
which we now turn.

\section{Higher Order Estimates}

      We have so far analyzed the chiral symmetry breaking phase transition
using the ladder gap equation, i.e. the SD equation with the lowest order
kernel,
and the running gauge coupling determined by the two-loop $\beta$
function. In order to consider higher order effects we first
develop
a gauge-invariant technique to estimate the critical coupling without
relying on the intricacies of the SD equation.

In Ref.  \cite{ALM}, it was noted that to lowest order the SD criticality
condition
can be written in the form
\beq
\gamma(2- \gamma) = 1 ~,
\label{crit}
\eeq
where $\gamma$ is the anomalous dimension of the quark mass operator.
To all orders in perturbation theory, this condition is gauge
invariant (since $\gamma$ is gauge invariant) and is equivalent to the
condition
\cite{rainbow}
$\gamma =1$ mentioned previously in the text.  However if these
conditions are truncated at  a finite order in  perturbation theory they
lead to different results. We will take Eq.~(\ref{crit}) to define the critical
coupling order by order, since it allows us to reproduce the
known leading order result.

 Through three loops $\gamma$ is given
in the $\overline{\rm MS}$ scheme by \cite{nikhef}
\beq
\gamma = \gamma_0 \alpha +  \gamma_1 \alpha^2 +
\gamma_2 \alpha^3  + ...
\eeq
where
\begin{eqnarray}
\gamma_0 &=&
{{3\,C_2(R)}\over {2\,\pi }} \\
\gamma_1 &=&
{{1 }\over
   {16\,{{\pi }^2}}}
[ 3\,{{C_2(R)}^2} - {{10\,C_2(R)\,N_f}\over 3} + {{97\,C_2(R)\,N}\over 3}
] \\
\gamma_2 &=&
{{1 }\over {64\,{{\pi }^3}}}[
129\,{{C_2(R)}^3} - {{70\,C_2(R)\,{N_f^2}}\over {27}} -
     {{129\,{{C_2(R)}^2}\,N}\over 2} +
     {{11413\,C_2(R)\,{{N}^2}}\over {54}} \nonumber \\
&& +
     C_2(R)\,N_f\,N\,\left( -{{556}\over {27}} - 48\,\zeta(3) \right)  +
     {{C_2(R)}^2}\,N_f\,\left( -46 + 48\,\zeta(3) \right)]
\end{eqnarray}
Inserting this result in Eq. (\ref{crit}) and truncating to one-loop we find
\beq
2 \gamma_0 \alpha = 1.
\eeq
Solving for $\alpha$ we find a one-loop estimate of the critical coupling
that agrees with standard result:
\beq
\alpha_c^{(1)}= {{ \pi }\over{3 \, C_2(R)}} =
{{2 \pi \,N}\over{3\left(N^2-1\right)}}~.
\eeq

At two-loops the critical condition is
\beq
2 \gamma_0 \alpha + 2 \gamma_1 \alpha^2 - \gamma_0^2 \alpha^2 =1.
\label{twoL}
\eeq
Solving for $\alpha$ we find a two-loop estimate of the critical coupling:
\beq
\alpha_c^{(2)}=  {{36\,\pi }\over {45\,C_2(R) - 97\,N + 10\,N_f}}
\pm {{{\sqrt{24}}\,\pi \,{\sqrt{9\,C_2(R) + 97\,N - 10\,N_f}} }\over
        {{\sqrt{C_2(R)}}\,\left( -45\,C_2(R) + 97\,N - 10\,N_f \right) }}
     ~.
\eeq
The $+$ sign gives the positive root.  We compare this with the one-loop
estimate
by taking N large and using the value $N_f \approx 4 N$ corresponding to
criticality:
\beq
\alpha_c^{(2)} \approx {{(  \sqrt{11808} -72 )\, \pi }\over{69 N}} \approx
{{1.67}\over{ N}}.
\eeq
Numerically it can be seen that the ${\cal O}(\alpha^2)$ terms in the
criticality condition, Eq. (\ref{twoL}),
evaluated
at $\alpha = \alpha_c^{(2)}$ are typically about 25\% to 30\% of the leading
term for
$N_f  \approx 4 N$. It can also be seen numerically that for for $N_f
\approx 4 N$
the four-loop term \cite{nikhef} in $\gamma$
is larger than the three-loop term, so it is not appropriate to go beyond two
loops in
this expansion for these values of $N_f$, and we should only use the
three-loop
term as an estimate of the error in our calculation.

 Through three-loops, the $\beta$ function is given by
\begin{displaymath}
\beta(\alpha)=-b\alpha^2-c\alpha^3-d\alpha^4
\end{displaymath}
where $b$ and $c$ are given by Eqs. (\ref{b}) and (\ref{c}), and in the MS
scheme,
\begin{eqnarray}
d  &=& {1\over 32\pi^2}  \left(  {2857 N^3-1415 N^2 N_f + 79 N(N_f)^2\over
54} -  {205N\over  18}C_2(R) N_f \right. \\
\nonumber
&& \,\,\,\,\,\, \,\,\,\,\,\, \,\,\,\,\,\, \,\,\,\,\,
+ \left. {{11}\over{9}}C_2(R) (N_f)^2+ C_2(R)^2 N_f \right)
\end{eqnarray}
Since the three-loop term is scheme dependent we cannot obtain a scheme
independent answer without going to the same order in $\beta$ and $\gamma$, so
we will only use the three-loop term for error estimates.

In Table 1 we list some numerical results. We have computed the value of
$N_f^c$
for $SU(N)$ gauge theories for values of  $N$
ranging form 2 to 10, showing the results at different orders in perturbation
theory.   In section  4 (using the leading order estimate of the critical
coupling)
it was shown that
$N_f^c$ goes like $4N$ for large $N$. We see that going to two loops in the
criticality condition produces a small shift in this relation. We also list the
estimated value of the critical coupling at one and two loops.
We see that even though the percentage shift of the value
of $N_f^c$
is small, the higher order terms of the beta function make a
significant contribution at the critical point. For $N_c$ between 3 and 10 we
estimate
that the error in $N_f^c$ at two-loops is about 12\% from the truncation of the
$\beta$ function and about 10\% from the truncation of $\gamma$, while for
$N_c=2$
the errors are somewhat larger, around 14\% from each.  It is important to
emphasize
that these are simply numerical estimates of the next to leading
contributions. Even at large N, there is no obvious small parameter here
leading to a controlled expansion. Thus the smallness of still higher order
terms is not guaranteed.

\begin{table}
\begin{center}
\begin{tabular}{|r|r|r|r|r|r|}
\hline
$N_c$ & $N_f^c$ (1,2) & $N_f^c$ (2,2) & $N_f^c$ (2,3) & $\alpha_c^{(1)}$
& $\alpha_c^{(2)}$  \\
\hline\hline

2 & 7.86 & 8.27 & 7.12 & 1.4 & 1.11 \\  \hline
3 & 11.9 & 12.4 & 10.9 & 0.785 & 0.595 \\ \hline
4 & 15.9 & 16.6 & 14.6 & 0.559 & 0.412 \\ \hline
5 & 20.0 & 20.8 & 18.3 & 0.436 & 0.317 \\ \hline
6 & 24.0 & 24.9 & 22. & 0.359 & 0.258  \\ \hline
7 & 28.0 & 29.1 & 25.7 & 0.305 & 0.218 \\ \hline
8 & 32.0 & 33.3 & 29.4 & 0.266 & 0.189 \\ \hline
9 & 36.0 & 37.4 & 33.1 & 0.236 & 0.166 \\ \hline
10 & 40.0 & 41.6 & 36.8 & 0.212 & 0.149 \\ \hline
\end{tabular}
\end{center}
\caption{Estimates of  $N_f^c$.  The two numbers in parentheses
give the order used in the critical condition on $\gamma$ and the
$\beta$ function.  The comparison of the (2,2) and (2,3) give an
estimate of the error in truncating the $\beta$ function at two-loops.}

\end{table}

\section{Summary and Conclusions}

  In this paper, we have explored features of the chiral phase transition
in $SU(N)$ gauge theories. We have argued that the transition takes place
at a relatively large value of $N_f$ ($N_f^c \approx 4N$) where the infrared
coupling is determined by a fixed point accessible in the loop
expansion of the $\beta$ function, and that the transition can be studied
using a ladder gap equation.  Our higher order estimates suggest that
the estimate of $N_f^c$ is good to about 20\%.
To phrase things in physical terms, the effect of the light quarks
is to screen the long range force, eventually disordering the
system and taking it to the symmetric phase. That the transition takes
place at a relatively large value of $N_f$ means that the quarks are
relatively ineffective at long range screening.

With an infrared fixed point governing the transition, the order parameter
vanishes in a characteristic exponential fashion and all physical scales
vanish in the same way. There is
no finite set of light degrees of freedom that can be identified to form an
effective, Landau-Ginzburg theory. In the symmetric phase ($N_f > N_f^c$) ,
no light degrees of freedom are formed as $N_f \rightarrow N_f^c$. Thus the
transition is continuous but not conventionally second order. The validity
of the approach is considered by estimating higher order terms in both the
$\beta$ function and the anomalous dimension of the mass operator.

  In Ref. \cite{AppSel}, it was noted that single instanton effects in a theory
with an infrared fixed point seem capable of triggering a chiral phase
transition at similarly large
values of $N_f/N$. A detailed computation was carried only out for an $SU(2)$
gauge theory but the analysis indicated that  this could be the case at larger
values of $N$ as well.

It is interesting to compare our results with the phase structure of
supersymmetric $SU(N)$ theories where exact results are available
\cite{Seiberg}. In such theories there is also a large range of $N_f$ where
the theory is asymptotically free and an infrared fixed point occurs.  A
transition
to a strongly coupled phase occurs at $N_{\rm f,SUSY}^c = 3 N/2$.  Thus it
seems
plausible that
infrared fixed points are fairly generic in asymptotically free gauge theories
with a large number of flavors.  One prominent difference between the
supersymmetric
and non-supersymmetric cases is that the strongly coupled phase
$N+1<N_f \le N_{\rm f,SUSY}^c$
does not have chiral symmetry breaking or confinement for
$N>3$.
However a class of supersymmetric chiral gauge theories  (with
antisymmetric tensor fields) have been found
\cite{FiveEasy} where
the theory does go from an infrared fixed  point to confinement upon the
removal of one flavor.

  The results of this paper can be contrasted with preliminary lattice work
\cite {Mawh} and the instanton liquid model \cite{VelShu} which
suggest that the chiral transition takes place at much smaller values of
$N_f$ contrary to earlier lattice results \cite{lattice}.
The transition would then be an intrinsically strong coupling
phenomenon inaccessible to the methods used here. The quarks would
have to be
much more effective at long range screening than indicated
by the gap equation, disordering the system even in the presence of a
strong, attractive long range force. Further work on all these approaches
will be required to help to resolve this difference.

\noindent{\bf Appendix - Gauge Invariance and Collinear Divergences}

  We first discuss the gauge dependence of the
quark-antiquark scattering amplitude used in Section 4 to demonstrate
the absence of light excitations in the symmetric phase. We will then discuss
the presence of collinear divergences in this amplitude.
To demonstrate gauge invariance to leading order, we follow the analysis of
\cite{gaugeinv}.
As was done before we will take the incoming (Euclidean) momentum of the
initial quark and
antiquark to be $q/2$, and  have
a non-zero momentum transfer by assigning outgoing momenta
$q/2 \pm p$ for the final quark
and antiquark.
The SD equation  in the scalar channel (and in a covariant gauge with gauge
parameter $\xi$) is:
\begin{eqnarray}
T(p,q) &=& {{g^2 \,Z_1^{2}(p,q)}\over{4 \alpha_\xi \,Z_3(p)}} \pi+
{{4\pi^2 \lambda Z_4(p,q) \, p^2}\over{\Lambda_*^2}} \nonumber \\
& & +{{\pi p^2} \over{\alpha_\xi}}
\int {{d^4k}\over{(2 \pi)^4}} {{g^2 \, Z_1^2(p,k)}\over{Z_3(p-k)\,(p-k)^2}}
\,{{T(k,q)}\over{ k^2 \,Z_2^2(k)}}
\nonumber \\
& &+
{{4  \pi^2 p^2}\over{\Lambda_*^2}} \int {{d^4k}\over{(2 \pi)^4}} \lambda
Z_4(p,k)
\,{{T(k,q)}\over{ k^2 \,Z_2^2(k)}}  ~.
\end{eqnarray}
The renormalization factors $Z_1$, $Z_2$, $Z_3$, and
$Z_4$
correspond to the gauge vertex, the quark
wavefunction,  the gauge boson wavefunction, and the four-quark vertex
respectively;
and
\beq
\alpha_\xi = {{\pi}\over{(3+\xi)C_2(R)}}~.
\eeq
Using the definition of the renormalized couplings
\beq
g_R(p,k) = {{g \,Z_1(p,k)}\over{\sqrt{Z_3(p-k) \, Z_2(k)Z_2(p)}}}
\eeq
\beq
\lambda_R(p,k) = {{\lambda \,Z_4(p,k)}\over{ Z_2(k)Z_2(p)}}\
\eeq
and the approximations
\beq
{{g^2_R(p,k)}\over{4 \pi}}
  \approx  {{g^2}\over{4 \pi}}
{{Z_1(\max(p,k))}\over{Z_3(\max(p,k))Z_2(k)Z_2(p)}}
\equiv \alpha(\max(p,k))
\eeq
and
\beq
\lambda_R(p,k) \approx \lambda{{Z_4(\max(p,k))}\over{Z_2(k)Z_2(p)}}
\equiv \lambda(\max(p,k))
\eeq
we can perform the angular integrations to obtain
\begin{eqnarray}
T(p,q) &=& {{\alpha(p)Z_2(p)Z_2(q)}\over{\alpha_\xi}} \pi^2 +
4\pi^2 \lambda(p) Z_2(p)Z_2(q)\,{{ p^2}\over{\Lambda_*^2}} \nonumber\\
& & +{{1} \over{4\alpha_\xi}}\left(
\int_{q^2}^{p^2} {{dk^2}\over{k^2}} \alpha(p) {{Z_2^2(p)}\over{Z_2^2(k)}}
\,T(k,q) +
\int_{p^2}^{\Lambda_*^2} {{dk^2}\over{k^2}} \alpha(k) \,T(k,q) \,{{
p^2}\over{k^2}} \right)
\nonumber \\
& &+
 \int_{q^2}^{p^2}  {{dk^2}\over{k^2}} \lambda(p)
{{Z_2^2(p)}\over{Z_2^2(k)}}\,T(k,q) \,{{p^2}\over{\Lambda_*^2}}
+
 \int_{p^2}^{\Lambda_*^2}  {{dk^2}\over{k^2}} \lambda(k)\,T(k,q) \,{{
p^2}\over{\Lambda_*^2}} ~.
\end{eqnarray}
In order to get a gauge invariant result, it is helpful to divide the
scattering
amplitude by
the gauge dependent normalization factors of the four quark legs, so we
introduce
\beq
\tilde T(p,q) = {{T(p,q)}\over{Z_2(p)\,Z_2(q)}} ~.
\eeq
We then have
\begin{eqnarray}
\tilde T(p,q) &=& {{\alpha_*}\over{\alpha_\xi}} \pi^2 +
4\pi^2 \lambda \,{{ p^2}\over{\Lambda_*^2}} \nonumber\\
& & +{{\alpha_*} \over{4\alpha_\xi}}\left(
\int_{q^2}^{p^2} {{dk^2}\over{k^2}}  {{Z_2(p)}\over{Z_2(k)}} \,\tilde T(k,q) +
\int_{p^2}^{\Lambda_*^2} {{dk^2}\over{k^2}}  {{Z_2(k)}\over{Z_2(p)}} \, \tilde
T(k,q) \,{{ p^2}\over{k^2}} \right)
\nonumber \\
& &+  \lambda \left(
 \int_{q^2}^{p^2}  {{dk^2}\over{k^2}}
{{Z_2(p)}\over{Z_2(k)}}\, \tilde T(k,q) \,{{p^2}\over{\Lambda_*^2}}
+
 \int_{p^2}^{\Lambda_*^2}  {{dk^2}\over{k^2}}
{{Z_2(k)}\over{Z_2(p)}}\, \tilde T(k,q) \,{{
p^2}\over{\Lambda_*^2}}  \right)~,
\label{xieqn}
\end{eqnarray}
where we have used the fact that $\alpha(p)$ approaches a fixed point for $p
\ll \Lambda$.  Here we will be satisfied with a result to leading order in
$\alpha_*$,
neglecting terms suppressed by $\alpha_*^2$, $\lambda^2$, and $\alpha_*
\lambda$.
With this approximation we can also neglect the running of $\lambda$.  This is
actually not a
bad approximation, since in the infrared $\lambda(p)$ approaches a fixed-point
given by equation (\ref{lambdatilde}).
Now the RG solution for the quark wavefunction renormalization
is:
\beq
Z_2(p) = \left({{\Lambda_*^2}\over{p^2}}\right)^\gamma ~,
\eeq
where
\beq
\gamma = {{\alpha_* C_2(R)\, \xi}\over{4 \pi}} +{\cal O}(\alpha_*^2)
\eeq

Next we substitute the form
\begin{equation}
\tilde T(p,q) = A
\left({p^2 \over \Lambda_*^2}\right)^{{1 \over 2} + {1 \over 2} \eta}
 +B
\left({p^2 \over \Lambda_*^2}\right)^{{1 \over 2} - {1 \over 2} \eta} ~,
\label{xisol}
\end{equation}
into equation (\ref{xieqn}).
Integrating this equation we see that to leading order in $\alpha_*$
the $\xi$ dependent terms take the form
\beq
{{\alpha_*}\over{4 \alpha_\xi
\left( {{1}\over{2}} -   {{1}\over{2}}  \eta + \gamma \right)
\left( {{1}\over{2}} +  {{1}\over{2}}  \eta + \gamma \right)}}
\approx  1 + {\cal O}(\alpha_*^2)
\eeq
So our solution for the scattering amplitude (equations (\ref{A}) and
(\ref{B})) and
the conclusion
that there are no light scalar degrees of freedom as one approaches the
critical point from the symmetric side
of the critical curve are
gauge invariant results to leading order.

We next discuss the collinear divergences present in $T(p,q)$.
Consider the differential cross-section for the scattering of the quark and
antiquark at ${\cal
O}(\alpha^3)$.  If the invariant amplitude at ${\cal
O}(\alpha^2)$ is given by
${\cal M}$, then from
equations (\ref{sol})-(\ref{B}) we have, to next-to-leading order,
\begin{equation}
|{\cal M}|^2\approx{{9 \pi^2 \alpha^2 C_2(R)^2}\over{p^4}} +{{27 \pi
\alpha^3 C_2(R)^3}\over{2p^4}}
\left(1+\ln\left({{p^2}\over{q^2}}\right)\right)~,
\label{NTLO}
\end{equation}
The differential cross section is:
\begin{equation}
d\sigma_0 =
 (2 \pi)^4 \delta^{(4)}(p_1+p_2-q_1-q_2) |{\cal M}|^2 {{d^3q_1}\over{(2 \pi)^3
2 E_1}}
{{d^3q_2}\over{(2 \pi)^3 2 E_2}} ~,
\end{equation}
which gives
\begin{equation}
{{d\sigma_0}\over{dq_1 d\Omega_1 dq_2 d\Omega_2}} =
{1\over{(2 \pi)^2}} \delta^{(4)}(p_1+p_2-q_1-q_2) |{\cal M}|^2  {{E_1
E_2}\over{4}}.
\end{equation}
This is not, however, a physically observable cross-section.
To obtain a physically observable cross-section we must combine this with
the differential cross-
section where a collinear gluon (with momentum $k$ and implicit summation on
the
gauge index a) is emitted:
\begin{equation}
d\sigma_{1g} =
 (2 \pi)^4 \delta^{(4)}(p_1+p_2-q_1-q_2) |{\cal M}^a|^2
{{d^3q_1}\over{(2 \pi)^3 2 E_1}}
{{d^3q_2}\over{(2 \pi)^3 2 E_2}} {{d^3k}\over{(2 \pi)^3 2 k}}~,
\end{equation}
A physical experiment cannot separately resolve the collinear gluon and
quark, so it is appropriate to
frame the discussion in terms of the momentum of the observed jet (we consider
first the case where $k$ is
approximately collinear with $q_2$, so $q_j = q_2 +k$).  Changing variables we
have
\begin{equation}
{{d\sigma_{1g}}\over{dq_1 d\Omega_1 dq_j d\Omega_j}} =
{1\over{(2 \pi)^2}} \delta^{(4)}(p_1+p_2-q_1-q_j)   {{E_1 E_j}\over{4}}
\int{{d^3k}\over{(2 \pi)^32 k}}{{(E_j-k)}\over{ E_j}}\overline{|{\cal
M}^a|}^2.
\label{diffcross}\end{equation}
Thus, to see the cancellation of the collinear divergence we must add $|{\cal
M}|^2$ to the final integral in
equation (\ref{diffcross}).

In order to project out the scalar channel of the gluon emission amplitude, we
must contract the amplitude
with $\delta_{\rho\lambda}/4$ and
$[\gamma^\alpha,\gamma^\beta]_{\sigma\tau}/16$, where $\rho$ and $\lambda$
($\sigma$ and $\tau$) are
the Dirac indices of the initial (final) quark and antiquark.  We then have
\begin{equation}
 {\cal M}^a=-{{i g^3 C_2(R)}\over{p^2q_j^2}} {3\over 4} \left(\epsilon^\alpha
q^\beta_j
-\epsilon^\beta q^\alpha_j\right) T^a~,
\end{equation}
where $\epsilon^\alpha$ is the gluon polarization vector.  Squaring and
summing over gluons and gluon
polarizations we have:
\begin{equation}
\overline{ |{\cal M}^a|}^2=-{{g^6 C_2(R)^3}\over{p^4q_j^2}} {27\over 8} ~.
\end{equation}
Putting the gluon on shell ($k^2=0$), and performing the integration (with the
requirement that the gluon
momentum $k$ be within a small cone of opening angle $\delta$ around the
quark momentum $q_2$) we
have:
\begin{eqnarray}
\int{{d^3k}\over{(2 \pi)^32 k}}{{(E_j-k)}\over{ E_j}}\overline{|{\cal
M}^a|}^2
& \approx & -{{27 g^6 C_2(R)^3}\over{8
p^2}}\int_0^{E_j}{{dk\,k^2(E_j-k)}\over{(2 \pi)^2
2k}}
\int_0^{\delta} {{\theta d\theta}\over{q_2^2+(E_j-k)k\theta^2}}\nonumber \\
&\approx& -{{27 \pi \alpha^3 C_2(R)^3}\over{4p^4}}
\ln\left({{E_j^2\delta^2}\over{q_2^2}}\right)~.
\end{eqnarray}
where we have only kept  terms which diverge as $q_2^2 \rightarrow 0$. When
combined with
the integration over the region of phase space corresponding to $k$ being
approximately collinear with
$q_1$, and setting $q_1=q_2 = q$, we see that these terms cancel with the
$\ln(q^2)$ dependence in equation
(\ref{NTLO}), as expected \cite{KLN}.

\noindent \medskip\centerline{\bf Acknowledgments}
\vskip 0.15 truein
We thank S. Chivukula and R. Mawhinney for helpful discussions.
We also thank the Aspen Center for Physics where part of this work was
completed. LCRW thanks Prof. K. Tennakone for hospitality and support at the
IFS and A. Herat for help with computations.
This work was supported in part by the Natural Sciences and
Engineering Research Council of Canada; the Texas National Research
Laboratory Commission  through an SSC fellowship and under contracts
\#RGFY93-278 and  \#RGFY93-272;  by the Department of Energy under
contracts \#DE-FG02-92ER40704 and \#DE-FG02-91ER40676,  \#DE-FG-02-84ER40153,
\# DE-AC03-76SF00098; and by the National Science
Foundation under grant PHY-95-14797.

\vfill\eject
\noindent \medskip\centerline{\bf Figure Caption}
Figure 1.  Numerical solution of the Schwinger-Dyson equation  with a
running coupling possessing an infrared fixed point.  Here $\Sigma_0$ is the
dynamical
mass and $r$ is the ratio of the fixed point coupling to the critical coupling.
\newpage
\begin{figure}
\epsfbox[120 210 420 660]{sigma.ps}
\end{figure}

\vskip 0.15 truein



\begin{thebibliography}{99}

\bibitem{GW}D.Gross and F.Wilczek, {\em Phys.Rev. }{\bf D8}(1973)3633.

\bibitem{CasJones}W.E.Caswell, {\em Phys.Rev.Lett.} {\bf 33}, (1974) 244;
 D.R.T.Jones, {\em Nucl.Phys.} {\bf B75}, (1974) 531.

\bibitem{BZ}T.Banks and A. Zaks, {\em Nucl. Phys.} {\bf B196} (1982) 189.

\bibitem{ATWFP}T. Appelquist, J. Terning, and L.C.R. Wijewardhana, {\em
Phys. Rev. Lett.} {\bf 77} (1996) 1214 hep-ph/9602385.

\bibitem{CJT}J.M. Cornwall, R. Jackiw, and E. Tomboulis, {\em Phys. Rev.} {\bf
D10} (1974) 2428.

\bibitem{ATW2+1}T. Appelquist, J. Terning, and L.C.R. Wijewardhana, {\em
Phys. Rev. Lett.} {\bf 75} (1995) 2081, hep-ph/9402320 .


\bibitem{QED3}T. Appelquist, D. Nash, and L.C.R. Wijewardhana,
{\em Phys. Rev. Lett.} {\bf 60} (1988) 2575;  D.  Nash {\em Phys. Rev. Lett.}
{\bf 62} (1989) 3024; T. Appelquist and D. Nash, {\em Phys. Rev. Lett.} {\bf
64} (1990) 721.

\bibitem{Nash} D.  Nash {\em Phys. Rev. Lett.}
{\bf 62} (1989) 3024; T. Appelquist and D. Nash, {\em Phys. Rev. Lett.} {\bf
64} (1990) 721.

\bibitem{Kogut}E. Dagotto, J.B. Kogut, and A.Kocic, {\em Phys. Rev. Lett.}
{\bf 62}, 1083 (1989); {\em Nucl. Phys.} {\bf B334}, 279 (1990).

\bibitem{postmodern}T. Appelquist, J. Terning, and
L.C.R. Wijewardhana, {\em Phys. Rev. Lett.} {\bf 79} (1997) 2767,
hep-ph/9706238.

\bibitem{LaneRam} K. Lane, M.V. Ramana {\em Phys. Rev.} {\bf D44}
(1991) 2678.

\bibitem{walking}
B. Holdom, {\em Phys. Lett.} {\bf B150} (1985) 301;
K. Yamawaki, M. Bando, and K. Matumoto, {\em Phys. Rev. Lett.}
{\bf 56} (1986) 1335;
T. Appelquist, D. Karabali, and L.C.R. Wijewardhana, {\em Phys. Rev.
Lett.} {\em 57} (1986) 957;
T. Appelquist and L.C.R. Wijewardhana, {\em Phys. Rev} {\bf D35} (1987) 774;
T. Appelquist and L.C.R. Wijewardhana, {\em Phys. Rev} {\bf D36} (1987) 568.

\bibitem{Lambert}R.M.Corless, G.H.Gonnet, D.E.G.Hare, D.J.Jeffrey and
D.E.Knuth, On the Lambert W Function", Advances in Computational
Mathematics,5(1996)329.

\bibitem{Karliner} This form of the solution to the two loop
renormalization group
equation in a gauge theory with an infrared fixed point has also been noted
by  E.Gardi and M.Karliner, hep-ph/9802218 and by E.
Gardi, G. Grunberg, and M. Karliner, hep-ph/9806462.



%\bibitem{largeN}G. t'Hooft {\em Nucl. Phys.} {\bf B72} (1980) 461;
%S. Coleman, {\em Aspects of Symmetry} (Cambridge University Press, Cambridge,
%1985).


\bibitem{colemanwitten}S. Coleman and E. Witten,
{\em Phys. Rev. Lett.} {\bf 45} (1980) 100.

\bibitem{FT2}R. Pisarski and F. Wilczek, {\em Phys. Rev.} {\bf D29} (1984) 338.

\bibitem{FT1}For a review see F.Wilczek, {\em Nucl.Phys.} {\bf A566}  (1994)
123c;
{\em Int. J. Mod. Phys.} {\bf D3} Suppl. (1994) 63.

\bibitem{rainbow}H. Georgi and A. Cohen, {\em Nucl. Phys.} {\bf B314} (1989) 7.

\bibitem{classicrefs}T. Maskawa and H. Nakajima, {\em Prog Theor. Phys}
{\bf 52}
(1974) 1326; {\bf 54} (1976) 860;
R. Fukuda and T. Kugo {\em Nucl. Phys.}  {\bf B117} (1976) 250;
P.I. Fomin and V.A. Miransky, {\em Phys. Lett.} {\bf B64} (1976) 166.
P.I.Fomin, V.P.Gusynin,and V.A.Miransky,Phys.Lett.B 78 (1978) 136;

\bibitem{SD}V.A. Miransky {\em Il Nuovo Cimento} {\bf 90A} (1985) 149;
V.A. Miransky and P.I. Fomin, {\em Sov. J. Part. Nucl.} {\bf 16} (1985) 203.


\bibitem{TABrazil}
T. Appelquist,  in {\em Particles and Fields
(Mexican School)} J.L. Lucio and A. Zepeda eds. (World Scientific, Singapore,
1992).


\bibitem{Bardeen}W. Bardeen, C.N. Lueng and S.T. Love, {\em Nucl. Phys.} {\bf
B273}
(1986) 649.

\bibitem{NJL}Y. Nambu and G. Jona-Lasinio, {\em Phys. Rev.} {\bf 122},
(1961) 345.


\bibitem{ATW1} T. Appelquist, J. Terning, and L.C.R. Wijewardhana,  {\em Phys.
Rev.} {\bf D44}  (1991) 871.

\bibitem{criticalhier}S. Chivukula, M. Golden, E.H. Simmons, {\em Phys. Rev.
Lett.} {\bf 70} (1993)1587, hep-ph/9210276.

\bibitem{colemanweinberg}S. Coleman and E. Weinberg, {\em Phys. Rev.} {\bf D7}
(1973) 1888.

\bibitem{Miransky}V. Miransky, {\em Int. J. Mod. Phys.} {\em A8} (1993) 135.

\bibitem{miryam}V.A. Miransky  and K. Yamawaki {\em Phys. Rev.} {\bf D55}
(1997) 5051, hep-th/9611142; erratum ibid. {\bf D56} (1997) 3768; V.A.
Miransky, hep-ph/9703413.


\bibitem{chivukula}S. Chivukula, {\em Phys. Rev.}
{\bf D55} (1997) 5238, hep-ph/9612267.


\bibitem{S}M. Golden and L. Randall, {\em Nucl. Phys.} {\bf B361} (1991) 3
(1991);
B. Holdom and J. Terning, {\em Phys. Lett} {\bf B247} (1990) 88;
M. Peskin and T. Takeuchi, {\em Phys. Rev. Lett.} {\bf 65} (1990) 964.


\bibitem{ALM}T. Appelquist, K. Lane, and U. Mahanta, {\em Phys. Rev. Lett.}
{\bf 61} (1988) 1553.



\bibitem{nikhef}J.A.M. Vermaseren, S.A. Larin, T. Van Ritbergen,
hep-ph/9703284.
%\bibitem{4-loop}T. Van Ritbergen, J.A.M. Vermaseren, and S.A. Larin,
%{\em Phys.Lett.} {\bf B400}  (1997) 379, hep-ph/9701390.




\bibitem{AppSel}T. Appelquist and S. Selipsky,
{\em Phys. Lett.} {\bf B400} (1997) 364, hep-ph/9702404.



\bibitem{Seiberg} For a recent review see K. Intriligator and N. Seiberg,
{\em Nucl. Phys. Proc. Suppl.} {\bf  45BC} (1996)1;
{\em Nucl. Phys. Proc. Suppl.} {\bf  55B} (1996) 200; hep-th/9509066.


\bibitem{FiveEasy}J. Terning, {\em Phys. Lett.} {\bf B422} (1998)149;
hep-th/9712167.


\bibitem{Mawh} D. Chen, R.D. Mawhinney,  {\em Nucl. Phys. Proc. Suppl.}
{\bf 53} (1997)216 hep-lat/9705029; R.D. Mawhinney, hep-lat/9705030;
see also Y. Iwasaki, K. Kanaya, S. Kaya, S. Sakai, T. Yoshie,
{\em Nucl. Phys. Proc.Suppl.} {\bf 53} (1997) 449, hep-lat/9608125.

\bibitem{VelShu}M. Velkovsky, E. Shuryak, hep-ph/9703345.


\bibitem{lattice}J. B. Kogut and D. K. Sinclair, {\em Nucl. Phys.} {\bf B295}
[FS21] (1988) 465, F. Brown et. al., {\em Phys. Rev.} {\bf D46} (1992) 5655.


 \bibitem{gaugeinv}T. Appelquist , U. Mahanta, D. Nash, and L.C.R.
Wijewardhana,
{\em Phys. Rev.} {\bf D43}(1991) 646.

\bibitem{KLN}T. Kinoshita, {\em J. Math Phys.} {\bf 3} (1962); T. D. Lee and
M. Nauenberg, {\em Phys. Rev.} {\bf 133}  (1964) B1549.




\end{thebibliography}
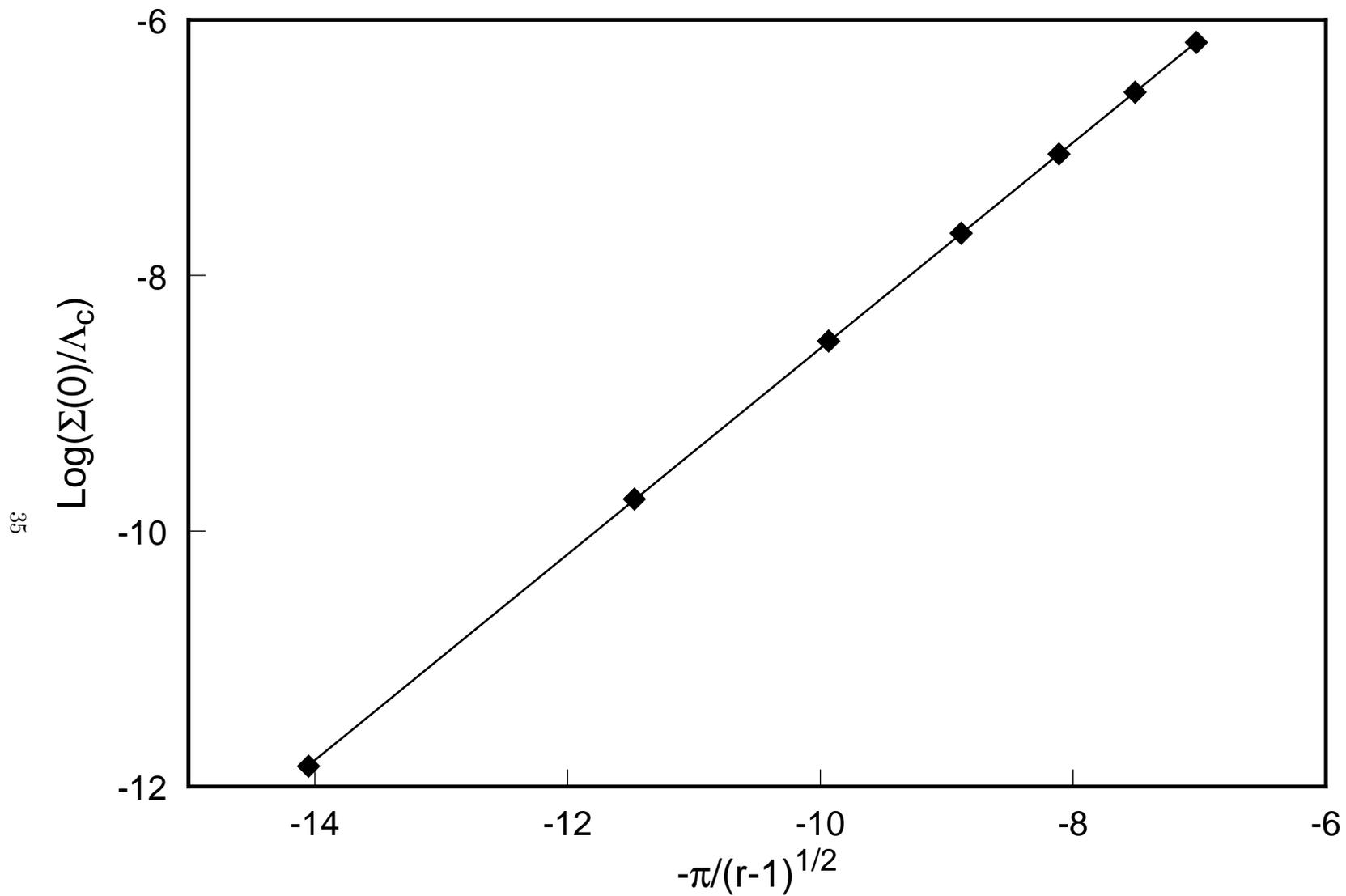
\end{document}